\documentclass[%
reprint,
superscriptaddress,
nofootinbib,
amsmath,amssymb,
aps,
prc,
floatfix,
]{revtex4-2}

\usepackage{graphicx}% Include figure files
\usepackage{dcolumn}% Align table columns on decimal point
\usepackage{bm}% bold math
\usepackage{color}
\usepackage{here}
\usepackage{hyperref}% add hypertext capabilities
\usepackage{nicefrac}
\usepackage{longtable}
\usepackage{lipsum}
\usepackage{ulem}
\usepackage{multirow}
\newcommand{\nuc}[2]{\hbox{$^{#1}$#2}}

\newcommand{\threesevensixsevenBe}{$3767(5)$} %1867+1900

\newcommand{\oneeightsixsevenBe}{$1866(3)$}
\newcommand{\oneninezerozeroBe}{$1900(4)$}
\newcommand{\twoonethreefiveBe}{$2131(4)$}
\newcommand{\threeonesixzeroBe}{$3159(5)$}

\newcommand{\medtwoplus}{$-75(3)$}
\newcommand{\medzero}{$-1218(5)$}
\newcommand{\medtwo}{$-473(5)$}

\begin{document}

%\preprint{APS/123-QED}

\title{In-beam $\gamma$-ray spectroscopy towards the proton dripline: The curious case of \nuc{32}{Ar}}
%\title{First indications for unexpectedly large mirror-energy differences in the \nuc{32}{Ar}-\nuc{32}{Si} pair}% Force line breaks with \\

%
\author{T.~Beck}
\email{beck@frib.msu.edu}
\altaffiliation[Present address: ]{LPC Caen, 14050 Caen Cedex, France}
\affiliation{Facility for Rare Isotope Beams, Michigan State University, East Lansing, Michigan 48824, USA}
\author{A.~Gade}
\affiliation{Facility for Rare Isotope Beams, Michigan State University, East Lansing, Michigan 48824, USA}
\affiliation{Department of Physics and Astronomy, Michigan State University, East Lansing, Michigan 48824, USA}
\author{B.A.~Brown}
\affiliation{Facility for Rare Isotope Beams, Michigan State University, East Lansing, Michigan 48824, USA}
\affiliation{Department of Physics and Astronomy, Michigan State University, East Lansing, Michigan 48824, USA}
%\author{J.A.~Tostevin}
%\affiliation{Department of Physics, Faculty of Engineering and Physical Sciences, University of Surrey, Guildford, Surrey GU2 7XH, United Kingdom}
%If included, put funding in acknowledgments, if not, thank Jeff there.
\author{Y. Utsuno}
\affiliation{Advanced Science Research Center, Japan Atomic Energy Agency, Tokai, Ibaraki 319-1195, Japan}
\affiliation{Center for Nuclear Study, University of Tokyo, Hongo, Bunkyo-ku, Tokyo 113-0033, Japan}
\author{D.~Weisshaar}
\affiliation{Facility for Rare Isotope Beams, Michigan State University, East Lansing, Michigan 48824, USA}
\author{D.~Bazin}
\affiliation{Facility for Rare Isotope Beams, Michigan State University, East Lansing, Michigan 48824, USA}
\affiliation{Department of Physics and Astronomy, Michigan State University, East Lansing, Michigan 48824, USA}
\author{K.W.~Brown}
\affiliation{Facility for Rare Isotope Beams, Michigan State University, East Lansing, Michigan 48824, USA}
\affiliation{Department of Chemistry, Michigan State University, East Lansing, Michigan 48824, USA}
\author{R.J.~Charity}
\affiliation{Department of Chemistry, Washington University, St. Louis, Missouri 63130 USA}
\author{P.J.~Farris}
\affiliation{Facility for Rare Isotope Beams, Michigan State University, East Lansing, Michigan 48824, USA}
\affiliation{Department of Physics and Astronomy, Michigan State University, East Lansing, Michigan 48824, USA}
\author{S.A.~Gillespie}
\affiliation{Facility for Rare Isotope Beams, Michigan State University, East Lansing, Michigan 48824, USA}
\author{A.M.~Hill}
\affiliation{Facility for Rare Isotope Beams, Michigan State University, East Lansing, Michigan 48824, USA}
\affiliation{Department of Physics and Astronomy, Michigan State University, East Lansing, Michigan 48824, USA}
\author{J.~Li}
\affiliation{Facility for Rare Isotope Beams, Michigan State University, East Lansing, Michigan 48824, USA}
\author{B.~Longfellow}
\altaffiliation[Present address: ]{Lawrence Livermore National Laboratory, Livermore, California 94550, USA}
\affiliation{Facility for Rare Isotope Beams, Michigan State University, East Lansing, Michigan 48824, USA}
\affiliation{Department of Physics and Astronomy, Michigan State University, East Lansing, Michigan 48824, USA}
\author{W.~Reviol}
\affiliation{Physics Division, Argonne National Laboratory, Argonne, Illinois 60439, USA}
\author{D.~Rhodes}
\altaffiliation[Present address: ]{Lawrence Livermore National Laboratory, Livermore, California 94550, USA}
%\altaffiliation[Present address: ]{TRIUMF, 4004 Wesbrook Mall, Vancouver, BC V6T 2A3, Canada}
\affiliation{Facility for Rare Isotope Beams, Michigan State University, East Lansing, Michigan 48824, USA}
\affiliation{Department of Physics and Astronomy, Michigan State University, East Lansing, Michigan 48824, USA}

\date{\today}% It is always \today, today, but any date may be explicitly specified

\begin{abstract}

High-resolution in-beam $\gamma$-ray spectroscopy was used to study excited states 
of the neutron-deficient nucleus \nuc{32}{Ar} populated in fast-beam induced 
four- and six-nucleon removal reactions from \nuc{36,38}{Ca}.
One new $\gamma$-ray transition and indications for an additional two were found, 
allowing for a glimpse at the level scheme beyond the $2^+_1$ state.
The nature of the new \oneninezerozeroBe-keV transition is discussed in the context of 
the known energy spectrum of the mirror nucleus \nuc{32}{Si}
and shell-model calculations using the FSU and SDPF-M cross-shell effective interactions.
Its resulting parent state at \threesevensixsevenBe\,keV, more than 1.3~MeV above the proton separation energy, 
is tentatively assigned to have mixed $sd$-shell and $2p$-$2h$ character. %and quantum numbers $J^{\pi}=0^+$. 
It might either be the mirror of the $J^{\pi}=2^+_2$ state of \nuc{32}{Si} at $4230.8(8)$\,keV,
but with a decay branch favoring a transition to the $2^+_1$ over the ground state,
or the mirror of the $4983.9(11)$-keV state with quantum numbers $0^+$.
The resulting mirror-energy differences of \medtwo~and \medzero\,keV are both sizable when compared to systematics; 
in the latter case it would, in fact, be among the largest reported to date in the nuclear chart
or suggest the potential existence of an additional, hitherto unidentified, low-lying $0^+$ state of \nuc{32}{Si}.
\end{abstract}

\pacs{Valid PACS appear here}% PACS, the Physics and Astronomy Classification Scheme.
%\keywords{Suggested keywords}%Use showkeys class option if keyword display desired
\maketitle

%%%%%%%%%%%%%%%%%%%%%%%%%%%%%%%%%%%%%%%%%%%%%%%%%%%%%%%%%%%%%%%
%	Introduction
%%%%%%%%%%%%%%%%%%%%%%%%%%%%%%%%%%%%%%%%%%%%%%%%%%%%%%%%%%%%%%%

%\section{Introduction}

%One of the most fundamental challenges of nuclear physics is the creation of a predictive model of all nuclei;
%ranging from light to heavy masses and from the valley of stability to the periphery of the nuclear chart.
%Precision experimental data enables an informed selection of model features and parameters
%while discrepancies between theoretical predictions and experiment expose model deficiencies.
%Undeniably, one of the most well-studied regions in the nuclear chart
%is the $sd$ shell, spanning from \nuc{16}{O} ($Z,N=8$) to \nuc{40}{Ca} ($Z,N=20$).
%It exhibits a 

Neutron-deficient nuclei, i.e. nuclei lighter than the corresponding $\beta$-stable isotopes
with at times larger number of protons $Z$ than neutrons $N$,
%found between the valley of stability and the proton dripline, 
are key for understanding nucleosynthesis mechanisms such as the $rp$ process~\cite{Wal81a}, 
might exhibit modes of proton radioactivity~\cite{Pfu12a},
and can provide a complementary view on nuclear forces~\cite{Hol13a}.
%Continuous progress at rare-isotope facilities~\cite{Blu13a} and in nuclear theory
%have the proton dripline experimentally established up to medium mass~\cite{Pfu23a} 
%and yielded {\it ab-initio} predictions of it up to iron~\cite{Str21a}, respectively.
As for the proton dripline up to medium-mass nuclei, the progress at rare-isotope facilities~\cite{Blu13a}
has allowed to establish its location~\cite{Pfu23a} and nuclear theory has yielded {\it ab-initio} predictions 
of it up to iron~\cite{Str21a}.
In the $sd$ shell, spanning from \nuc{16}{O} ($Z,N=8$) to \nuc{40}{Ca} ($Z,N=20$),
spectroscopic information on excited states of many neutron-deficient isotopes up to 
or even beyond the dripline is available.
Recent experimental highlights from this region include
the observation and first spectroscopy of the three-proton emitter \nuc{31}{K},
located four neutrons beyond the dripline~\cite{Kos19b}, or the spectroscopy of \nuc{35}{Ca} 
potentially suggesting doubly-magic character of \nuc{36}{Ca}~\cite{Lal23a}.

The present work is focussed on the neutron-deficient \nuc{32}{Ar} ($Z=18$, $N=14$) nucleus.
It is the lightest proton-bound, even-even argon isotope as
two-proton decay was observed for the ground-state of \nuc{30}{Ar}~\cite{Muk15a}.
%{\color{red}If it comes out in time, we can also refer to this new mass measurement of \nuc{31}{Ar}:
%\href{https://journals.aps.org/prl/accepted/17079Y52G521889481fc9192e61acc620960715f3}{Link}}
The low-lying structure of its astrophysically interesting, heavier isotopic neighbors \nuc{33,34}{Ar} 
is known from combined particle-$\gamma$ spectroscopy~\cite{Gad04a,Cle04a,Ken20a}.
Previous studies involving \nuc{32}{Ar} mainly concerned its mass~\cite{Bla03a} and 
$\beta$-decay properties~\cite[and references therein]{Bla21a} for the search of physics beyond the standard model.
Removal of a strongly-bound neutron from \nuc{32}{Ar}~\cite{Gad04b} triggered the discovery of the strong correlation
between the experimental-to-theoretical cross-section ratio, $R_s$, and the asymmetry of proton and neutron binding
in heavy-ion-induced nucleon-removal reactions~\cite{Gad08a,Tos14a,Tos21a}.
Information on its excited states, on the other hand, remained sparse.
Hitherto, only the $2^+_1$ state of \nuc{32}{Ar} at $1867(8)$\,keV~\cite{NDS32} has been identified experimentally
using scintillator arrays~\cite{Cot02a,Bue07a} and segmented germanium detectors~\cite{Yon06a}.
Excited states above the comparatively low proton separation energy of \mbox{$S_p($\nuc{32}{Ar}$)=2455(4)$\,keV}~\cite{Wan21a},
revealing the proximity of \nuc{32}{Ar} to the proton continuum, remained unknown.
The present work provides new information on excitations beyond the $2^+_1$ state,
populated in the \nuc{197}{Au}(\nuc{36}{Ca},\nuc{32}{Ar}$+\gamma$)$X$ and 
\nuc{9}{Be}(\nuc{38}{Ca},\nuc{32}{Ar}$+\gamma$)$X$ multi-nucleon removal reactions,
from first high-resolution in-beam $\gamma$-ray spectroscopy using a modern tracking array.

%%%%%%%%%%%%%%%%%%%%%%%%%%%%%%%%%%%%%%%%%%%%%%%%%%%%%%%%%%%%%%%
%	Experiment and results
%%%%%%%%%%%%%%%%%%%%%%%%%%%%%%%%%%%%%%%%%%%%%%%%%%%%%%%%%%%%%%%

%\section{Experiment and Results}
%\label{sec:res}

The reaction channels discussed here share the same secondary beams of unstable \nuc{36,38}{Ca} nuclei
as the experiments reported in Refs.~\cite{Dro23a} and \cite{Gad20a,Gad22a,Gad22b,Bec23a,Bec24a}, respectively.
They were produced through fragmentation of a stable \nuc{40}{Ca} beam,
accelerated to 140\,MeV/u by the Coupled Cyclotron Facility of the National Superconducting Cyclotron Laboratory~\cite{Gad16a}, 
on a $799$-mg/cm$^2$ \nuc{9}{Be} production target.
The fragments were subsequently separated in the A1900 fragment separator~\cite{Mor03a}, 
using a $300$\,mg/cm$^2$ aluminum wedge degrader and
with the momentum acceptance restricted to $\Delta p/p=0.25\%$.
The resulting beams, which for \nuc{36}{Ca} and \nuc{38}{Ca} had purities of $11$~\cite{Dro23a} and $85$\%,
were impinged on $257$ and $188$\,mg/cm$^2$-thick \nuc{197}{Au} and \nuc{9}{Be} secondary reaction targets, respectively,
located at the target position of the S800 magnetic spectrograph~\cite{Baz03a}.
Incoming species were identified from their time-of-flight differences taken between two plastic scintillators 
at the exit of the A1900 and the object position of the S800 analysis beam line.
Outgoing projectile-like reaction products were selected on an event-by-event basis in the S800 focal plane~\cite{Yur99a} using time-of-flight differences taken between two plastic scintillators at the object position of the analysis beam line 
and the S800 focal plane and their energy loss measured in the spectrograph's ionization chamber.

\begin{figure}[t]
\centering
\includegraphics[trim=0 0 0 0,width=1.025\linewidth,clip]{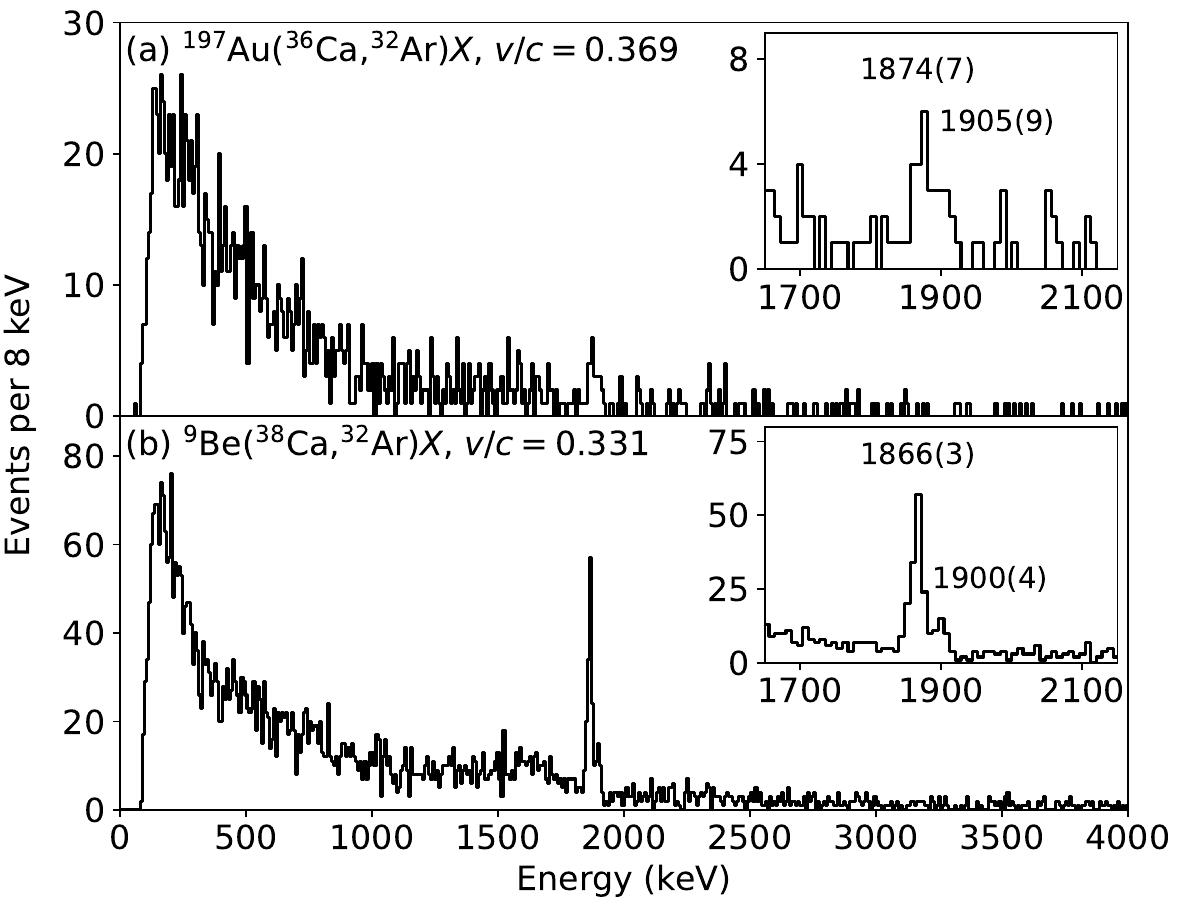}
\caption[]{Doppler-reconstructed $\gamma$-ray spectra of \nuc{32}{Ar} produced in the
(a) \nuc{197}{Au}(\nuc{36}{Ca},\nuc{32}{Ar}$+\gamma$)$X$ and 
(b) \nuc{9}{Be}(\nuc{38}{Ca},\nuc{32}{Ar}$+\gamma$)$X$ reactions.
The insets show the region around the $2^+_1\to0^+_1$ transition, 
unveiling the existence of a new $\gamma$-ray transition at \oneninezerozeroBe\,keV.
The difference in resolution between both reactions can be explained by the strongly
differing stopping in the gold and beryllium targets.}
\label{fig:spec}
\end{figure}

Prompt $\gamma$ rays emitted in flight by the excited reaction products were detected with the high-resolution tracking array 
GRETINA~\cite{Pas13a,Wei17a}, comprising twelve detector modules of four 36-fold segmented germanium crystals each.
Employing online signal decomposition, the spatial coordinates of the  $\gamma$-ray interaction points with the highest energy deposition in GRETINA
are identified and used for event-by-event Doppler correction, furthermore incorporating information on the momentum vector 
of \nuc{32}{Ar} in the exit channel, exploiting the track-reconstruction capabilities of the S800 spectrograph.
Resulting $\gamma$-ray spectra of \nuc{32}{Ar} from both production pathways are displayed in Fig.~\ref{fig:spec}.
The spectrum [Fig.~\ref{fig:spec}\,(b)] produced in the \nuc{9}{Be}(\nuc{38}{Ca},\nuc{32}{Ar})$X$ reaction 
features peaks at \oneeightsixsevenBe~and \oneninezerozeroBe\,keV.
The former is attributed to the ground-state transition of the $2^+_1$ state at $1867(8)$\,keV~\cite{NDS32}
and corroborates the results of Refs.~\cite{Yon06a,Bue07a}, which report a higher excitation energy than Cottle \textit{et al.}~\cite{Cot02a}. 
The $\gamma$-ray transition at \oneninezerozeroBe\,keV is unknown until now.
Both peaks can also be identified in the $\gamma$-ray spectra [Fig.~\ref{fig:spec}\,(a)] 
following the \nuc{197}{Au}(\nuc{36}{Ca},\nuc{32}{Ar})$X$ reaction, though with lower statistics.
The different intensity ratios \mbox{$I_{\gamma}(1900)/I_{\gamma}(1866)=54_{-29}^{+46}$} and $17(7)\%$
obtained in the \nuc{36}{Ca}- and \nuc{38}{Ca}-induced reactions, respectively, might be traced to  differing population characteristics of the four- and six-nucleon removal reactions~\cite{Obe06a,Lon23a}.
Also, the $\gamma$-ray spectrum of \nuc{32}{Ar} produced in the \nuc{9}{Be}(\nuc{37}{Ca},\nuc{32}{Ar})$X$ reaction, 
which is shown in Ref.~\cite{Bue07a}, suggests the existence of a second peak at around $1.9$\,MeV, 
although not explicitly discussed in that publication.

\begin{figure}[t]
\centering
\includegraphics[trim=0 0 0 0,width=1.025\linewidth,clip]{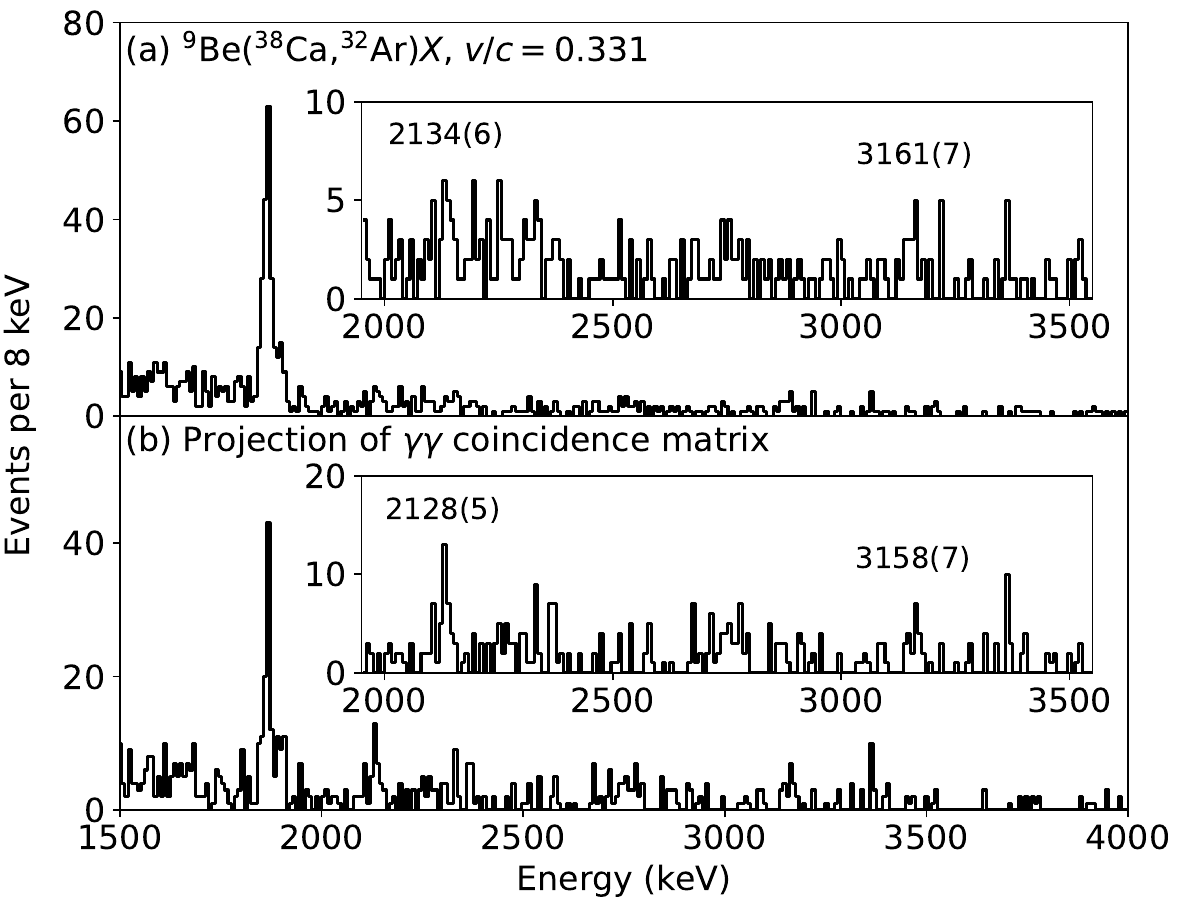}
\caption[]{(a) Doppler-reconstructed $\gamma$-ray spectrum of \nuc{32}{Ar} produced in the 
\nuc{9}{Be}(\nuc{38}{Ca},\nuc{32}{Ar}$+\gamma$)$X$ reaction and using nearest-neighbor add-back~\cite{Wei17a}.
(b) Projection of the $\gamma\gamma$ coincidence matrix from the same reaction.
The insets enlarge the high-energy part of the spectrum with low statistics and suggest the existence
of two candidate $\gamma$-ray transitions of \nuc{32}{Ar}.}
\label{fig:gg}
\end{figure}

Limited statistics do not allow for the use of $\gamma\gamma$ coincidences to clarify the relation between both $\gamma$ rays.
Figure~\ref{fig:gg}\,(b) shows the total projection of the $\gamma\gamma$ coincidence matrix
with the two peaks identified in the $\gamma$-ray singles spectrum clearly visible.
The spectrum in coincidence with the most prominent $2^+_1\to0^+_1$ transition
has at most one count in any bin above 1\,MeV.
However, the projected coincidence matrix reveals the likely existence of two additional, previously unknown
$\gamma$-ray transitions.
They are also identified in the $\gamma$-ray singles spectra with nearest-neighbor add-back~\cite{Wei17a} included.
The weighted averages of their energies extracted from the different spectra shown in Fig.~\ref{fig:gg}\,(a) and (b)
are \twoonethreefiveBe~and \threeonesixzeroBe\,keV.
%

%%%%%%%%%%%%%%%%%%%%%%%%%%%%%%%%%%%%%%%%%%%%%%%%%%%%%%%%%%%%%%%
%	Discussion
%%%%%%%%%%%%%%%%%%%%%%%%%%%%%%%%%%%%%%%%%%%%%%%%%%%%%%%%%%%%%%%

%\section{Discussion}

Since it is not possible to determine whether the \oneeightsixsevenBe~and \oneninezerozeroBe-keV $\gamma$ rays
form a cascade, two possible placements for the latter need to be considered.
The new $\gamma$ ray either corresponds to the ground-state transition of an excited state of \nuc{32}{Ar} at \oneninezerozeroBe\,keV
or the decay of a proton-unbound state at \threesevensixsevenBe\,keV into the $2^+_1$ level.
Guidance can be obtained from the level scheme of the mirror nucleus \nuc{32}{Si}~\cite{NDS32}.
Its $2^+_1$ state is found at $1941.4(3)$\,keV~\cite{NDS32}; this corresponds to a conventional  
mirror-energy difference (MED) of \medtwoplus\,keV for the \nuc{32}{Ar}/\nuc{32}{Si} pair.
%In the calculations the $2^+_1$ state is of $sd$-shell origin and its excitation energy is reproduced closely.
There is no state in close proximity to it; the $2^+_2$ state is found more than $2$\,MeV higher in energy.
Based on this observation, the existence of a  \oneninezerozeroBe-keV state of \nuc{32}{Ar} can be refuted
since it would correspond to an MED of more than $2.3$\,MeV.
This is well beyond the range normally attributed to Coulomb-interaction induced isospin breaking~\cite{Hen20a}
and even exceeds the ``colossal'' MED of the $0^+_2$ state in the \nuc{36}{Ca}/\nuc{36}{S} mirror pair~\cite{Lal22a} 
by a factor of four.
It can, instead, be concluded that the \oneninezerozeroBe-keV transition most likely stems
from the $\gamma$ decay of a proton-unbound state at \threesevensixsevenBe\,keV into the $2^+_1$ level.
%The second excited state of \nuc{32}{Si} is the $2^+_2$ state at $4230.8(8)$\,keV~\cite{NDS32}
%which predominantly decays into the ground state.
%%It can be ruled out that the former is the $2^+_2$ state, which for \nuc{32}{Si} predominantly decays into the ground state.
%Using the known branching ratio \mbox{$I_{\gamma}(2^+_2\to2^+_1)/I_{\gamma}(2^+_2\to0^+_1)=0.61(5)$}~\cite{NDS32},
%a peak containing $22(8)$ counts against an average background of one count per bin
%should be clearly identifiable at \threesevensixsevenBe\,keV in the spectrum of Fig.~\ref{fig:spec}, panel~(b).

In order to clarify the nature of the new \threesevensixsevenBe-keV state of \nuc{32}{Ar},
further input from theory is needed.
Since the lowest-lying known negative-parity states of \nuc{32}{Si} are found at around $5$\,MeV,
the level structure is first compared to shell-model calculations carried out with the code NuShellX~\cite{Bro14a} 
and using the Florida State University (FSU) cross-shell effective interaction~\cite{Lub19a,Lub20a} 
in the extensive $spsdf\!p$ model space.
%As shown in Fig.~\ref{fig:fsu}, the level scheme calculated for $A=32$ is in good agreement 
%with the available experimental data for \nuc{32}{Si}.
The level scheme calculated for $A=32$ is in good agreement 
with the available experimental data for \nuc{32}{Si}.
%
%\begin{figure}[t]
%\centering
%\includegraphics[trim=20 350 100 60,width=1.025\linewidth,clip]{fig3.pdf}
%\caption[]{{\color{red}@BAB: Can we please put here a new figure with three panels and $sd$-shell states in green.
%For \nuc{32}{Ar} with the $2^+_1$ and the tentative $0^+/2^+_2$ state at \threesevensixsevenBe\,keV
%as well as $S_p(\nuc{32}{Ar})$?}
%Comparison of excitation energies obtained from shell-model calculations with the
%FSU cross-shell excitations (b) to the mirror nuclei (a) \nuc{32}{Ar} and (c) \nuc{32}{Si}.
%The length of the level markers indicates the angular-momentum quantum number and the color 
%$sd$-shell origin $\Delta=0$ (green), negative parity $\Delta=1$ (blue), and positive parity $\Delta=2$ (red).
%States with unknown quantum numbers are marked with black dots.
%}
%\label{fig:fsu}
%\end{figure}
\begin{figure}[t]
\centering
\includegraphics[trim=0 0 0 0,width=1.025\linewidth,clip]{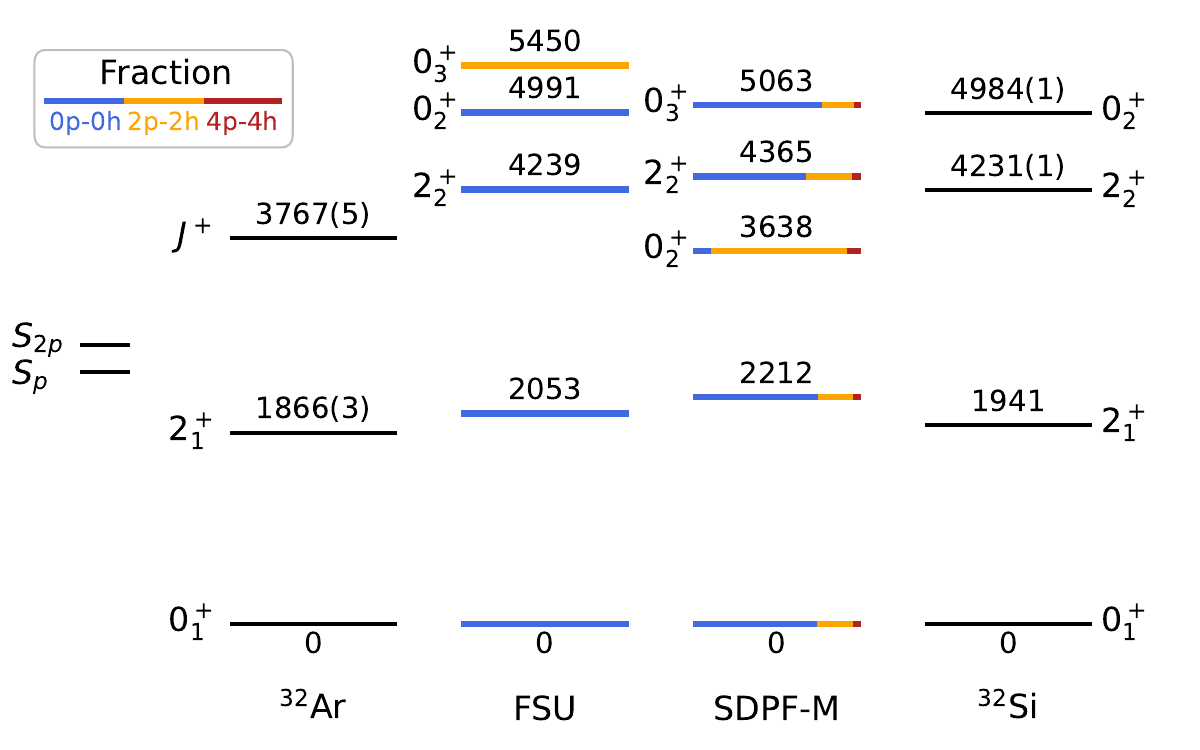}
\caption[]{Comparison of excitation energies in keV obtained from shell-model calculations with the
FSU and SDPF-M cross-shell excitations to the mirror nuclei \nuc{32}{Ar} and \nuc{32}{Si}.
The lines marking levels obtained from the shell model are color-coded by the fractions
of $0p$-$0h$ (blue), $2p$-$2h$ (orange), and $4p$-$4h$ (red)  configurations in their wave functions.
Energies and quantum numbers of excited states of \nuc{32}{Si} and proton-separation energies of \nuc{32}{Ar} 
are taken from Refs.~\cite{NDS32} and \cite{Wan21a}, respectively. 
%Angular-momentum and parity quantum numbers in parentheses are tentative.
}
\label{fig:theo}
\end{figure}
The structure of the calculated states can be characterized by $\Delta$
which signifies the number of nucleons promoted from the $sd$ to the $pf$ shell. 
In this notation, states with $\Delta=0$ are of $sd$-shell origin and those with $\Delta=1$ have negative parity and cross-shell nature.
The $0^+_1$ and $2^+_{1,2}$ states for $A=32$, for instance, are of $sd$-shell origin.
%and the latter's excitation energy is reproduced exceptionally well by the shell model.
The $3^-_1$ and $5^-_1$ states found at $5288.8(8)$ and $5581(4)$\,keV~\cite{NDS32} in \nuc{32}{Si},
are predicted at $5897$ and $6003$\,keV, respectively.
Above \mbox{$S_p($\nuc{32}{Ar}$)=2455(4)$\,keV}~\cite{Wan21a}, all  states with $\Delta=0$ and $1$ 
calculated in the shell model are predicted to exclusively proton-decay to final states of \nuc{31}{Cl}.
Thus, the conjectured state at \threesevensixsevenBe\,keV must be attributed
to a $\Delta=2$ configuration in the FSU spectrum.
%furthermore corroborating the exclusion of a $2^+_2$ assignment.
%states with $\Delta=3$ are only found above \red{$X$\,MeV}.
Since $\gamma$-ray transitions are possible only between states differing by at most one unit in $\Delta$,
the wave function must also feature components with $\Delta=0$.
The delicate interplay of these two configurations can produce an excited state
which has at least comparable proton- and $\gamma$-decay probabilities, 
with the latter reaching the $sd$-shell $2^+_1$ state.
Due to the low $Q$ value -- \mbox{$S_{2p}($\nuc{32}{Ar}$)=2719.0(18)$\,keV}~\cite{Wan21a} --
the probability for $2p$ decay of this state is likely negligible~\cite{Gri03a}.
Since the FSU cross-shell effective interaction deals in pure $\Delta$ configurations, 
it does not yield information on an admixed state.
Its properties can instead be studied using the SDPF-M cross-shell effective interaction~\cite{Uts99a,Uts01a}
which allows for the mixing of different particle-hole configurations between the $sd$ and $pf$ shells.
With the model space truncated to excitations with at most $\Delta=4$, the calculations, 
which were performed with the code KSHELL~\cite{Shi19b}, yield dominant $sd$-shell nature 
for the $0^+_1$ and $2^+_{1,2}$ states in agreement with the FSU results.
An intruder $0^+$ state, characterized by an $84\%$ $2p$-$2h$ component in its wave function,
is predicted at $3638$\,keV and is absent from the FSU spectrum in this energy region.
The next higher $0^+_3$ state ($80\%$ $0p$-$0h$) is found at $5063$\,keV, 
very close to the $\Delta=0$ $0^+_2$ state placed at $4991$\,keV 
in the calculations using the FSU effective interaction.
A comparison of %the available experimental data for low-lying states of \nuc{32}{Ar} and \nuc{32}{Si} to
the energy levels predicted by the FSU and SDPF-M calculations is shown in Fig.~\ref{fig:theo}.

Experimental information on $2p$-$2h$ states is available in the vicinity of the \nuc{32}{Ar}/\nuc{32}{Si} mirror pair.
%\red{What about \nuc{34}{Ar}? $0^+$ states are known at $3873(3)$, $4967(4)$, and $5909(12)$\,keV. Is one $\Delta=2$?}
In \nuc{34}{Si}, the $\Delta=2$ $0^+$ state is found at $2719(3)$\,keV~\cite{Rot12a}, 
in agreement with shell-model calculations using the FSU effective interaction~\cite{Bro22a}.
In neighboring \nuc{32}{Si}, a $0^+$ state is found at $4983.9(11)$\,keV~\cite{NDS32},
which decays via $\gamma$-ray emission to the $2^+_1$ state with a half life of $T_{1/2}<0.3$\,ps~\cite{Pro72a}.
Its population in the \nuc{30}{Si}$(t,p)$ reaction is reported to be too strong for a pure $sd$-shell nature, and 
DWBA calculations using $sd$-shell transfer amplitudes~\cite{Chu77a} fail to adequately describe its angular distribution~\cite{For82a}.
Fortune \textit{et al.} noted that a small $(sd)^{-2}(f_{7/2})^2$ admixture of less than $20\%$ to the state's wave function 
suffices to reproduce the measured cross section~\cite{For82a}.
This finding, along with the state's excitation energy, fits exceptionally well to the $0^+_3$ state 
from the SDPF-M spectrum which contains a $20\%$ $2p$-$2h$ contribution to its wave function.
%This state's energy is reasonably close to the shell-model prediction
%which places the pure $\Delta=2$ $0^+$ state at around 5.4\,MeV.
%In the calculations, the next available $\Delta=0$ $0^+$ state above the ground state
%for the discussed mixing scenario is found at $5.0$\,MeV.
%
\begin{table}[t]
\caption{Compilation of single-particle proton-decay half-lives $T_{1/2}^{sp}(p)$ and 
spectroscopic factors $C^2S$ from the FSU and SDPF-M shell-model effective interactions.
Comparison to $\gamma$-decay half-lives $T_{1/2}$ measured for \nuc{32}{Si}~\cite{NDS32}
reveals the predominance of the proton-decay channel to final states of \nuc{31}{Cl} 
for proton-unbound states of \nuc{32}{Ar} with $\Delta=0$ or $1$.
}
\label{tab:pdecay}
\begin{ruledtabular}
\begin{tabular}{cccllcc}
\multicolumn{3}{c}{Single-particle proton decay}		&\multicolumn{2}{c}{$C^2S$}	&$J^{\pi}_i$	&\nuc{32}{Si}~\cite{NDS32}\\
$J^{\pi}_f$		&$(n,l,j)$		&$T_{1/2}^{sp}(p)~(ps)$	&FSU		&SDPF-M		&			&$T_{1/2}~(ps)$\\\colrule
$3/2^+$		&$1d_{3/2}$	&					&$1.98$		&$1.50$		&$0^+_1$		&\\
$3/2^+$		&$2s_{1/2}$	&					&$0.011$		&$0.0011$	&$2^+_1$		&$0.78(22)$\\
\\
$3/2^+$		&$1d_{3/2}$	&$1.4\times10^{-6}$		&$0.0076$	&$0.18$$^1$	&$0^+_2$		&$<0.30$\\
$3/2^+$		&$1d_{3/2}$	&$1.4\times10^{-6}$		&$0$$^2$		&$0.010$		&$0^+_3$		&\\
$3/2^+$		&$2s_{1/2}$	&$3.0\times10^{-8}$		&$0.082$		&$0.021$		&$2^+_2$		&$0.26(9)$\\
$3/2^+$		&$2p_{3/2}$	&$7.2\times10^{-8}$		&$0.050$		&			&$3^-_1$		&$0.15(35)$\\
$3/2^+$		&$1f_{7/2}$	&$1.7\times10^{-5}$		&$0.77$		&			&$5^-_1$		&$27(2)\times10^3$\\
\\
$1/2^+$ 		&$2s_{1/2}$   	&					&$1.26$		&$1.18$		&$0^+_1$		&\\  
$1/2^+$		&$2s_{1/2}$   	&$3.1\times10^{-5}$		&$0.31$		&$0.0021$$^1$	&$0^+_2$		&$<0.30$\\
$1/2^+$		&$2s_{1/2}$   	&$3.1\times10^{-5}$		&$0$$^2$		&$0.11$		&$0^+_3$		&\\
\end{tabular}
\end{ruledtabular}
$^1$ This state has $84\%$ $\Delta=2$.\\
$^2$ This state has $\Delta=2$.
\end{table}

As evident from Fig.~\ref{fig:theo}, two potential assignments for the new
\threesevensixsevenBe-keV state of \nuc{32}{Ar} must be considered.
It may be identified as the intruder $0^+_2$ state predicted in the shell-model calculations 
using the SDPF-M effective interaction which yields a close reproduction of the excitation energy.
However, this state is predicted to predominantly proton decay to the $3/2^+$ and $(1/2^+)$ states of \nuc{31}{Cl}
with $Q$ values of $1.312$ and $0.575$\,MeV, respectively,
as evident from Table~\ref{tab:pdecay}, which collects single-particle proton-decay half-lives $T_{1/2}^{sp}(p)$ and 
spectroscopic factors $C^2S$ obtained from the FSU and SDPF-M shell-model effective interactions.
With $T_{1/2}(p)=T_{1/2}^{sp}(p)/C^2S$, the spectroscopic factors must be less than about $10^{-6}$ 
to allow for $\gamma$ and proton decays to compete.
The calculated values for the states discussed here, however, are much larger.
The small spectroscopic factor of $0.0076$ for the proton decay of the $0^+_2$ state to the
$3/2^+$ ground state of \nuc{31}{Cl} could be made arbitrarily close to zero
by a small mixing with the $0^+_1$ state.
But this mixing will not significantly change the larger spectroscopic factor of $C^2S=0.31$ for its
decay to the first-excited $(1/2^+)$ state of  \nuc{31}{Cl},
maintaining the predominance of proton over $\gamma$ decay.
A comparison to the level scheme of \nuc{32}{Si} would make the \threesevensixsevenBe-keV state of \nuc{32}{Ar} 
the mirror of the $4983.9(11)$-keV $0^+$ state of \nuc{32}{Si}~\cite{NDS32}, 
resulting in a mirror-energy difference of \medzero\,keV.
This would be among the largest MED ever suggested, 
surpassing even the $0^+_2$ state of the \nuc{36}{Ca}/\nuc{36}{S} mirror pair~\cite{Lal22a}
and large continuum-driven energy differences for the $A=13$ and $16$ pairs
\nuc{13}{N}/\nuc{13}{C}~\cite{Ehr51a,Tho52a} and \nuc{16}{F}/\nuc{16}{N}~\cite{Ste14a}, respectively,
and rivaling the MEDs put forth for \nuc{24}{Si}/\nuc{24}{Ne}~\cite{Lon20a} and \nuc{14}{O}/\nuc{14}{C}~\cite{Cha19a}.
However, the dominant $sd$-shell character of the known $0^+_2$ state of \nuc{32}{Si}~\cite{For82a}
might suggest the existence of an additional, until now unidentified, lower-lying $0^+$ state of \nuc{32}{Si}.
In the \nuc{30}{Si}$(t,p)$ experiments of Fortune \textit{et al.}~\cite{For82a}, its signal might have been obscured 
by reactions on target contaminants in the energy region expected for a normal-sized MED without weak-binding effects~\cite{Hen20a}.
In the vicinity of the \nuc{32}{Ar}/\nuc{32}{Si} mirror pair, $0^+_2$ states are indeed found in this energy region;
at $3667.4(8)$\,keV~\cite{NDS30} for the $N=14$ isotone \nuc{30}{S}, $3873(3)$\,keV~\cite{NDS34} for \nuc{34}{Ar},
and at $3787.72(5)$\,keV~\cite{NDS30} for \nuc{30}{Si}.

Assuming instead that the \threesevensixsevenBe-keV state is the mirror 
of the $2^+_2$ state of \nuc{32}{Si} at $4230.8(8)$\,keV~\cite{NDS32}
yields an MED of \medtwo\,keV which is still large when compared to systematics~\cite{Hen20a}
but significantly smaller than for the $0^+$ assignment discussed above.
In \nuc{32}{Si}, the $2^+_2$ state decays predominantly to the ground state
in contrast to the shell-model calculations with the FSU and SDPF-M effective interactions
which predict strong transitions to the $2^+_1$ and $0^+_2$ states, respectively.
Using the known branching ratio \mbox{$I_{\gamma}(2^+_2\to2^+_1)/I_{\gamma}(2^+_2\to0^+_1)=0.61(5)$}~\cite{NDS32},
a peak containing $22(8)$ counts against an average background of one count per bin
should be clearly identifiable at \threesevensixsevenBe\,keV in the spectrum of Fig.~\ref{fig:spec}\,(b).
Small admixtures from energetically close-lying states to the $2^+_2$ state might, 
reminiscent of the situation encountered for some rare-earth nuclei~\cite{Zil90a,Bec20a},
have altered its decay characteristics to feature a dominant transition to the ground state of \nuc{32}{Si}.
%It is unlikely that an energetically low-lying, excited state of \nuc{32}{Si} is missing from the literature, 
%given the very good agreement between experiment and theory in this energy region. 

Lastly, it has to be examined if a state with partial $2p$-$2h$ nature can be populated in 
the \nuc{197}{Au}(\nuc{36}{Ca},\nuc{32}{Ar})$X$ and \nuc{9}{Be}(\nuc{38}{Ca},\nuc{32}{Ar})$X$ reactions discussed here.
The weakening of the \mbox{$Z=20$} shell gap in the region of the neutron-deficient calcium isotopes, 
which has recently been further corroborated through invariant-mass spectroscopy of \nuc{34}{K} and \nuc{37,38}{Sc}~\cite{Dro24a}, 
leads to an increased proton $pf$-shell occupancy in the ground states of \nuc{36,38}{Ca}~\cite{Dro23a,Bec23a}.
Using the ZBM2 shell-model effective interaction~\cite{Cau01a}, fractions of $31$, $28$ and $25\%$
are found for the $2p$-$2h$ proton intruder components in the ground-state wave functions of \nuc{36,37,38}{Ca}, respectively.
Hence, few-nucleon removal reactions from these configurations, as employed here, 
may indeed populate final states of \nuc{32}{Ar} with two protons in the $pf$ shell.
This might furthermore explain why the \threesevensixsevenBe-keV state appears to have also been populated 
in the \nuc{9}{Be}(\nuc{37}{Ca},\nuc{32}{Ar})$X$ reaction~\cite{Bue07a}.
For two-neutron removal from \nuc{34}{Ar}, on the other hand, population cross sections can be obtained
from a combination of eikonal reaction dynamics and SDPF-M two-nucleon amplitudes~\cite{Tos04a,Tos06a}.
These calculations predict a vanishing population probability for the intruder $0^+_2$ state and
comparable cross sections for the $2^+_1$ and $2^+_2$ states.
In the experiment of Yoneda \textit{et al.} only the ground-state decay of the $2^+_1$ state has been observed~\cite{Yon06a}
whereas a proton decay of the $2^+_2$ state, as expected from the half-lives given in Table~\ref{tab:pdecay}, 
would have remained undetected.

%%%%%%%%%%%%%%%%%%%%%%%%%%%%%%%%%%%%%%%%%%%%%%%%%%%%%%%%%%%%%%%
%	Summary
%%%%%%%%%%%%%%%%%%%%%%%%%%%%%%%%%%%%%%%%%%%%%%%%%%%%%%%%%%%%%%%

%\section{Summary}

In summary, excited states of the neutron-deficient, near-dripline nucleus \nuc{32}{Ar} populated in 
four- and six-nucleon removal reactions from \nuc{36,38}{Ca} ions at intermediate energy were studied.
From subsequent high-resolution $\gamma$-ray spectroscopy, a previously unknown \oneninezerozeroBe-keV 
$\gamma$-ray of \nuc{32}{Ar} is identified. 
It is interpreted as the decay of a new excited state at \threesevensixsevenBe\,keV into the $2^+_1$ state.
Shell-model calculations using the FSU cross-shell effective interaction suggest that it must be of mixed 
$0p$-$0h$ and $2p$-$2h$ nature to enable the observed $\gamma$ decay to the $sd$-shell $2^+_1$ state.
From comparison to the mirror nucleus \nuc{32}{Si}, two potential mirror states are identified;
the $2^+_2$ state, which requires the mixed character of the state in \nuc{32}{Ar} to significantly alter its decay properties,
a vanishing population cross section in $2n$-removal in contrast to theoretical predictions,
and a large mirror-energy difference, or the $0^+_2$ state, resulting in one of the largest MED ever reported.
Calculations employing the SDPF-M shell-model effective interaction predict an intruder $0^+$ state in the right energy range,
potentially suggesting that its counterpart in the \nuc{32}{Si} level scheme is hitherto unobserved.
For $\gamma$ decay to compete, it would, however, require a reduction of the spectroscopic factor
for proton decay to the $(1/2^+)$ state of \nuc{31}{Cl} by a factor of roughly $10^6$.
Indications of additional, unplaced $\gamma$-ray transitions of \nuc{32}{Ar} at \twoonethreefiveBe~and \threeonesixzeroBe\,keV 
are found but further experimental studies are needed to corroborate their existence and 
also to confirm the placement of the \oneninezerozeroBe-keV $\gamma$ ray.
Any chosen reaction mechanism has to facilitate a population of excited nuclear states with two protons in the $pf$ shell. %, 
%ruling out the application of direct one- or two-neutron removal reactions~\cite{Han03a,Tos04a,Tos06a} to study its wave function
%assuming this cross-shell parentage is not contained in the knockout parent and
%on the basis of cross-section calculations using SDPF-M two-nucleon amplitudes.
Inelastic proton scattering in inverse kinematics~\cite{Gad08b}, for example,
which has been shown to efficiently populate off-yrast states~\cite{Ril19a},
might help to shed light on the nature of the \threesevensixsevenBe-keV state.
Conversely, the identification of a $0^+$ state of \nuc{32}{Si} at around $3.7$\,MeV,
for instance in two-neutron transfer or $\beta$-decay experiments,
has the potential to support a $0^+$ assignment for the \threesevensixsevenBe-keV state of \nuc{32}{Ar}.

%%%%%%%%%%%%%%%%%%%%%%%%%%%%%%%%%%%%%%%%%%%%%%%%%%%%%%%%%%%%%%%
%	Acknowledgments
%%%%%%%%%%%%%%%%%%%%%%%%%%%%%%%%%%%%%%%%%%%%%%%%%%%%%%%%%%%%%%%

\begin{acknowledgments}

The authors thank J.~A. Tostevin for providing theoretical predictions of two-neutron removal cross sections.
Discussions with W. Nazarewicz, A. Volya, and S. Wang are acknowledged.
This work was supported by the U.S. NSF under Grants No. PHY-1565546 and No. PHY-2110365, 
by the DOE NNSA through the NSSC, under Award No. DE-NA0003180, 
and by the U.S. DOE, Office of Science, Office of Nuclear Physics, 
under Grants No. DE-SC0020451 \& DE-SC0023633 (MSU) and No. DE-FG02-87ER-40316 (WashU) 
and under Contract No. DE-AC02-06CH11357 (ANL).
GRETINA was funded by the DOE, Office of Science. 
Operation of the array at NSCL was supported by the DOE under Grant No. DE-SC0019034.

\end{acknowledgments}

\bibliography{bib_32Ar}% Produces the bibliography via BibTeX.

\end{document}